\documentclass[a4paper]{article}
\usepackage{amssymb}
\usepackage{amsmath}

\def\be{\begin{equation}}
\def\eea{\end{eqnarray}}
\def\bea{\begin{eqnarray}}
\def\ee{\end{equation}}
\def\d{\mathrm{d}}
\def\p{\varphi}
\def\s{\sigma}

\author{Masoud Alimohammadi\footnote{alimohmd@ut.ac.ir}
\\  {\small Department of Physics, University of Tehran,}
\\ {\small North Karegar Ave., Tehran, Iran.}}
\title{ Asymptotic behavior of $\omega$ in general quintom model}
\date{}
\begin{document}
\maketitle
\begin{abstract}
For the quintom models with arbitrary potential $V=V(\p,\s)$, the
asymptotic value of equation of state parameter $\omega$ is
obtained by a new method. In this method, $\omega$ of stable
attractors are calculated by using the ratio $\d\ln V/\d \ln a$ in
asymptotic region. All the known results, have been obtained by
other methods, are reproduced by this method as specific examples.
\end{abstract}

\section{Introduction}
The recent observations on type Ia supernova \cite{1}, large scale
structure \cite{2} and cosmic microwave background radiation
\cite{3}, support the claim that our universe has accelerating
expansion and it is composed of dark energy ($73\%$), dark matter
($23\%$) and baryonic matter ($4\%$).

The simplest candidate for dark energy is a cosmological constant
$\Lambda$ of order ($10^{-3}$ev)$^4$, but it suffers from
conceptual problems such as fine-tuning and coincidence problems
\cite{4}. As the alternative to cosmological constant, the
dynamical models have been introduced.

One of the important parameters of dark energy is the equation of
state parameter $\omega =p/\rho$, which must satisfy $\omega<-1/3$
if one interested in accelerating universe. The time dependence of
$\omega$, i.e. $\omega(z)$, is usually used to explain the
dynamical nature of our universe. Some astrophysical data seems to
slightly favor an evolving dark energy and shows a recent
$\omega=-1$, the so-called phantom-divide-line, crossing \cite{5}.

Many of dynamical dark energy models are based on scalar fields.
The simplest of these models is the quintessence model which
consists of one normal scalar field $\p$ \cite{6}. In all
quintessence models, always $\omega>-1$. The other kind of scalar
field model is phantom model. In this model, there exists a
phantom scalar field $\s$ which its kinetic term appears with
minus sign. In phantom model, $\omega$ always satisfies
$\omega<-1$ \cite{7}. None of these two models can explain the
$\omega=-1$ crossing and at least two scalar fields are needed, in
models known as hybrid models, for occurrence of this transition.
One of these models is the quintom model which consists of one
quintessence and one phantom fields \cite{8}. One can show that in
all quintom models with slowly-varying potentials, the transition
from $\omega>-1$ to $\omega<-1$ is always possible \cite{9}.
Related to $\omega=-1$ crossing, it must be added that the
occurrence of this transition is also possible in a model with
only one scalar field, but this field must be coupled to
background matter field \cite{10}.

One of the interesting topics in dark energy problem is the
studying of the late-time, or asymptotic, behavior of physical
quantities. In this connection, the investigation of attractors of
the various dark energy models is one of the main tools \cite{11}.
Among the attractors of a dynamical system, the stable attractors
determine the late-time behavior of various quantities, including
the equation of state parameter $\omega$. Despite its usefulness,
the attractors' studies are restricted to specific potentials.
This is because the choosing of the set of convenient variables,
which results in an autonomous set of equations, is a highly
nontrivial task and depends on the functional form of the
considered potential. Therefore seeking other methods for studying
the late-time behavior of, for instance, $\omega$ is an important
task.

In ref.\cite{12}, the asymptotic behavior of $\omega_\s$ of
phantom models has been investigated for potentials which
$V(\s)\rightarrow\infty$ asymptotically. There, it has been shown
that all such models are divided into three classes, characterized
by $\omega_\s\rightarrow -1$, $\omega_\s\rightarrow \omega_0<-1$
and $\omega_\s\rightarrow -\infty$. In ref.\cite{13}, the
evolution of a special quintom model, the so-called hessence
model, has been discussed in $\omega-\omega'$ plane. The hessence
model is a quintom model with potential $V(\p,\s)=V(\p^2-\s^2)$.
$\p$ and $\s$ denote the quintessence and phantom fields,
respectively. In this model, there is, effectively, only one
dynamical field $\phi=\sqrt{\p^2-\s^2}$. The late time behavior of
this model has been classified in $\omega-\omega'$ plane in
\cite{13} and it has been shown that there exist four distinct
regions in this plane. Both of the above mentioned studies are
restricted to cases with only one dynamical dark energy field.

In this paper we want to study the asymptotic behavior of equation
of state parameter $\omega$ for quintom models with arbitrary
potential. Since it is a two-field-component model, there exists
new features which do not appear in single-component models of
\cite{12} and \cite{13}. It is must be added that the attractors
of quintom models with two classes of potentials has been recently
studied in \cite{14-0}. The potentials are restricted to cases in
which the scalar fields do not diverge, and the potentials which
behave asymptotically as an exponential potential. As we will see,
there are too many examples that do not belong to the above
mentioned cases, (to be specific, the examples B, C, D, F and G of
section 2), and one must seek another method, as we do, to obtain
the asymptotic behavior of the system.

The paper is organized as follows. In section two, we briefly
review the quintom model and derive a relation which describes the
dynamical evolution of the quintom potential. This equation is one
that we use to obtain the asymptotic behavior of equation of state
parameter $\omega$ of an arbitrary quintom model. It is the same
relation which exists in hessence model \cite{13}, and reduces to
the known relations in quintessence \cite{14} and phantom
\cite{15} models in appropriate limits. In section three, we use
this relation to obtain the asymptotic value of $\omega$ for
several examples. The examples are divided to two classes: the
potentials which have no quintessence-phantom interaction, i.e.
$V(\p,\s)=V_1(\p)+V_2(\s)$, and the potentials containing the
interaction term $V_{\rm int}(\p,\s)$. In all examples, it is seen
that our results coincide with those obtained by other methods,
e.g. the attractors' study or the numerical study of fields'
equations.

We use the units $c=\hbar=8\pi G=1$ throughout the paper.

\section{ The evolution equation of quintom potential}
we consider a spatially flat Friedman-Robertson-Walker space-time
in comoving coordinates $(t,x,y,z)$
 \be\label{1}
 \d s^2=-\d t^2+a^2(t)(\d x^2+\d y^2+\d z^2),
 \ee
where $a(t)$ is the scale factor. The quintom dark energy consists
of the quintessence field $\p$ and the phantom field $\s$, with
Lagrangian
 \be\label{2}
 {\cal L}_{\rm de}=\frac{1}{2} \partial_\mu\p \partial^\mu\p
 -\frac{1}{2}\partial_\mu\s \partial^\mu\s -V(\p,\s).
 \ee
The energy density $\rho_{\rm de}$ and pressure $p_{\rm de}$ of
the homogenous quintom dark energy are
 \begin{align}\label{3}
 \rho_{\rm de}=&\frac{1}{2}{\dot \p}^2 - \frac{1}{2}{\dot \s}^2
 +V(\p,\s),\nonumber\\
  p_{\rm de}=&\frac{1}{2}{\dot \p}^2 - \frac{1}{2}{\dot \s}^2
 -V(\p,\s),
 \end{align}
 and the evolution equations of the fields are
   \be\label{4}
   {\ddot \p}+3H{\dot \p} +\frac{\partial V(\p,\s)}{\partial \p}=0,
   \ee
and
  \be\label{5}
   {\ddot \s}+3H{\dot \s} -\frac{\partial V(\p,\s)}{\partial \s}=0.
   \ee
Here $H(t)={\dot a(t)}/a(t)$ is the Hubble parameter and 'dot'
denotes the time derivative. The Friedman equations are
  \be\label{6}
   3H^2=\rho_{\rm de},
   \ee
and
  \be\label{7}
   -2{\dot H}=\rho_{\rm de}+p_{\rm de}.
   \ee
Note that the equations (\ref{4})-(\ref{7}) are not independent
and eq.(\ref{7}) is obtained from eqs.(\ref{4})-(\ref{6}). The
equation of state parameter $\omega=p_{\rm de}/\rho_{\rm de}$ is
therefore
  \be\label{8}
  \omega=\frac{{\dot \p}^2 -{\dot \s}^2 -2V(\p,\s)}
  {{\dot \p}^2 -{\dot \s}^2 +2V(\p,\s)}.
  \ee
Because of the field equations (\ref{4}) and (\ref{5}), the
quintessence field $\p$ rolls down the potential to reach a
minimum of $V$, while the phantom field $\s$ falls up the
potential $V$ and settles in a maximum of $V$. So the late-time
values of $(\p,\s)$ are the saddle point of potential $V$, we
denote it by $(\p^*,\s^*)$. We will use this fact in our analysis.

To obtain the evolution equation of potential $V$, we first define
the function
  \be\label{9}
  x=\left\vert\frac{1+\omega}{1-\omega}\right\vert=\frac{\pm ({\dot \p}^2- {\dot \s}^2)}
  {2V},
  \ee
if $\omega\neq -1$. Differentiating $\ln x$ with respect to $\ln
a$ results in
  \be\label{10}
  \frac{\d\ln x}{\d\ln a}=\frac{1}{H} \left\{ \frac
  {2({\dot \p}{\ddot \p}-{\dot \s}{\ddot \s})}
  {{\dot \p}^2- {\dot \s}^2} - \frac {\dot V} {V} \right\}.
  \ee
Using eqs.(\ref{4}) and (\ref{5}) and noting that ${\dot V}=
(\partial V/ \partial \p){\dot \p} + (\partial V/ \partial
\s){\dot \s}$, eq.(\ref{10}) can be written as
  \be\label{11}
  \frac{\d\ln x}{\d\ln a}=\frac{1}{H} \left\{ -6H-\frac{2}{1+\omega }
   \frac {\dot V} {V} \right\},
  \ee
or
  \be\label{12}
  -\frac {\dot V} {V}=3H(1+\omega )\left( 1+\frac{1}{6}
  \frac{\d\ln x}{\d\ln a} \right).
  \ee
This is a general equation which can be used for $\p =0$ case
(which the quintom model becomes phantom model), $\s =0$ case
(where the model is the quintessence model), and when
$V(\p,\s)=V(\phi=\sqrt{\p^2-\s^2})$, which the quintom model
becomes hessence model. In all above mentioned cases, the
potential is a function of only one field variable, e.g. the field
$\theta$, and therefore eq.(\ref{12}) can be reduced to an
equation for the field-derivative of potential. This is because
${\dot V}=(\d V/\d\theta ){\dot \theta}$. But in the quintom
model, it is not the case.

It is more suitable to write the left-hand-side of eq.(\ref{12})
in terms of derivative with respect to $\ln a$. The result is
  \be\label{13}
  -\frac {V'} {V}=3(1+\omega )\left( 1+\frac{1}{6}
  \frac{x'}{x} \right),
  \ee
in which
  \be\label{14}
  f':=\frac{\d f}{\d\ln a}=\frac {1}{H}{\dot f}.
  \ee
We are going to use eq.(\ref{13}) to determine the late-time
behavior of $\omega$ for quintom models, i.e. to obtain the value
$\omega_0$ for stable attractors of quintom models. To do so, we
must first evaluate the ratio $x'/x$ for attractor solutions. The
first step in studying the attractors is to introduce a set of
convenient dimensionless variables, which in study of quintom
models, three of these variables are
  \be\label{15}
  x_\p=\frac{\dot\p}{\sqrt{6}H}\,\,\, , \,\,\,
  x_\s=\frac{\dot\s}{\sqrt{6}H}\,\,\, , \,\,\,
  y=\frac{\sqrt{V}}{\sqrt{3}H}.
  \ee
The other variables depend on the functional form of the
potential, but the above three variables always exist. In terms of
these variables, the ratio $x'/x$ becomes
  \be\label{16}
  \frac{x'}{x}=\frac{2\omega'}{1-\omega^2}= \frac {2\left[ x_\p y^2x_\p'-
  x_\s y^2x_\s' + (x_\s^2-x_\p^2)y y'\right]} {y^2(x_\p^2 -x_\s^2)}.
  \ee
For attractor solutions, one has $x_\p'=x_\s'=y'=0$, so
$x'/x\rightarrow 0$. Of course one must be more careful for
situation in which $y^2(x_\p^2 -x_\s^2)=0$. In quintom models, the
fields $\p$ and $\s$ satisfy different equations of motion, i.e.
$\p$ moves toward the minimum of potential while $\s$ goes to the
maximum of $V$. Therefore we expect that in stable attractor
solutions, $y^2\sim V(\p^*,\s^*)\neq 0$ and usually $x_\p^2\neq
x_\s^2$. In next section, we see that these conditions are
fulfilled in all known examples, except the example F which we
discuss it separately. So for stable attractors, the equation of
state parameter $\omega$ tends to $\omega_0$, obtained from
  \be\label{17}
  \omega\rightarrow\omega_0=-1-\frac{1}{3}\frac{V'}{V}.
  \ee

\section{Examples}
In this section we apply the main result (\ref{17}) to situations
which have been studied by other methods. These examples are
divided to two classes: The quintom models without
quintessence-phantom interaction and the potentials contain this
interaction term. We discuss these two kinds of potential
separately.
\subsection { The potentials $V(\p,\s)=V_1(\p)+V_2(\s)$}
For potentials of the form $V(\p,\s)=V_1(\p)+V_2(\s)$, the
asymptotic configurations of the fields $\p$ and $\s$ are
$(\p^*,\s^*)$, where $V_1(\p)$ becomes minimum at $\p^*$ and
$V_2(\s)$ attains its maximum at $\s^*$. In following we consider
four examples.\\ \\
 {\bf A. $V=V_{\p_0}e^{-\lambda_1\p}+
V_{\s_0}e^{-\lambda_2\s}$}\\ \\
 The attractors of this potential have been studied in \cite{16}.
 As the exponential function has no extremum points, so we can
 only use the fact that $\p$ always rolls down the potential and
 $\s$ falls up it. So $\p^*\rightarrow\infty$ and $\s^*\rightarrow
 -\infty$, from which $V(\p,\s)\rightarrow V_2(\s)$. The ratio
 $V'/V$ then becomes
  \be\label{18}
  \frac{V'}{V}\simeq\frac{V_2'}{V_2}= \frac{(\partial V_2/\partial
  \s)\s'}{V_2}=-\lambda_2\s'.
  \ee
Eq.(\ref{17}) then results in
  \be\label{19}
  \omega\rightarrow\omega_0=-1+\frac{1}{3}\lambda_2\s'.
  \ee
This is exactly the result obtained in \cite{16}. The only stable
attractor of this potential is characterized by $x_\p=0$,
$x_\s=-\lambda_2/\sqrt{6}$ and $y^2=1+\lambda_2^2/6$. So
  \be\label{20}
  \omega=\frac{x_\p^2-x_\s^2-y^2}{x_\p^2-x_\s^2+y^2} = -1
  -\frac{\lambda_2^2}{3},
  \ee
which is equal to (\ref{19}) ( note that $x_\s=\s'/\sqrt{6} =
-\lambda_2/\sqrt{6}$, results in $\s'=-\lambda_2$, so
$\omega=-1-\lambda_2^2/3=-1+(1/3)\lambda_2\s')$.\\ \\
{\bf B. $V=V_{\p_0}\p^\alpha+ V_{\s_0}\s^\alpha$ $(\alpha >0)$}\\ \\
The time-variation of $\omega$ of this potential has been studied
for two special cases $\alpha =2$ and $\alpha =1.8$ in \cite{161}.
$V_1(\p)$ is minimum at $\p^*=0$, but $V_2(\s)$ has no maximum,
although it is clear that $|\s^*|\rightarrow\infty$. So
$V(\p,\s)\rightarrow V_2(\s)$ and
  \be\label{21}
  \frac{V'}{V}\simeq\frac{V_2'}{V_2}=\left.
  \frac{2\s'}{\s}\right|_{\s\rightarrow\pm\infty}\longrightarrow
  0,
  \ee
from which eq.(\ref{17}) results in
  \be\label{22}
  \omega\rightarrow\omega_0=-1.
  \ee
This is the same result obtained for special cases in \cite{161}. It can be shown that
eq.(\ref{22}) is also valid for odd $\alpha$'s where $\varphi^*\rightarrow - \infty$.\\ \\
{\bf C. $V=V_{\p_0}e^{-\lambda_1\p^2}+
V_{\s_0}e^{-\lambda_2\s^2}$}\\ \\
$V_1(\p)$ has no minimum, but it is obvious that
$|\p^*|\rightarrow\infty$. $V_2(\s)$ is maximum at $\s^*=0$. So
again $V(\p,\s)\rightarrow V_2(\s)$, from which
  \be\label{23}
  \frac{V'}{V}\simeq\frac{V_2'}{V_2}=\left.
  -2\lambda_2\s\s'\right|_{\s=0}=0.
  \ee
This results in $\omega_0=-1$ which is consistent with numerical
calculation of \cite{16}.\\ \\
{\bf D. $V=V_0e^{\lambda\s^\alpha} \ (\alpha>0)$}\\ \\
Now we consider a phantom model with the above potential. In this
case $|\s^*|\rightarrow\infty$. Therefore
 \be\label{240}
   \frac{V'}{V}=\left.\alpha\lambda\s^{\alpha-1}\s'
   \right|_{\s=\pm\infty}=\begin{cases}
  0,& \alpha<1\\ \lambda\s',& \alpha=1\\ \infty,& \alpha>1\end{cases},
 \end{equation}
from which
  \be\label{250}
   \omega_0=\begin{cases}
  -1,& \alpha<1\\ -1-\lambda\s'/3,& \alpha=1\\ -\infty,&
  \alpha>1\end{cases}.
 \end{equation}
In above we use the fact that in $\alpha$=even cases, we have
$\s^*\rightarrow\pm\infty$. For $\s^*\rightarrow +\infty$, since
$\s'>0$, one has $\s^{\alpha-1}\s'\rightarrow +\infty$. For
$\s^*\rightarrow -\infty$, we know that $\s'<0$ and therefore
again $\s^{\alpha-1}\s'\rightarrow +\infty$. This general result
agrees with the numerical calculations of \cite{12} which have
been done for special cases $\alpha=0.5,1,$ and 2 with
$\lambda=1$. It is clear that eq.({\ref{250}) can be also used as
the late-time behavior of $\omega$ of a quintom model with the
potential
$V=V_{\p_0}e^{\lambda_1\p^\alpha}+V_{\s_0}e^{\lambda_2\s^\alpha}$.

\subsection{ The interacting potentials}
We now consider the potentials of the form
$V(\p,\s)=V_1(\p)+V_2(\s)+V_{\rm int.}(\p,\s)$, through three
following examples.\\ \\
{\bf E. $V=V_0e^{-\sqrt{6}(m\p+n\s)}$}\\ \\
The attractors of this quintom potential have been discussed in
\cite{17}. This potential has no extremum point, but it is clear
that $\p^*\rightarrow\infty$ and $\s^*\rightarrow -\infty$. The
ratio $V'/V$ is
  \be\label{24}
  \frac{V'}{V}=\frac{1}{V}\left( \frac{\partial V}{\partial \p}\p'
  +\frac{\partial V}{\partial \s}\s' \right) =-\sqrt{6} (m\p'
  +n\s'),
  \ee
so
  \begin{align}\label{25}
  \omega_0=&-1+\frac{\sqrt{6}}{3}(m\p'+n\s')\nonumber\\
   =&-1+2(mx_\p+nx_\s).
   \end{align}
In \cite{17}, it has been found two stable attractors, denoted
there by P and T solutions. In P-case, the coordinates are
$x_\p=m$, $x_\s=-n$ and $y^2=1-m^2+n^2\neq 0$, which result in
$\omega_{\rm P}=-1+2(m^2-n^2)$. This is the same as our result
(\ref{25}). In T-case, it has been obtained
$x_\p=m/(2(m^2-n^2)),\,x_\s=-n/(2(m^2-n^2))$ and
$y^2=1/(4(m^2-n^2))$, with $\omega_{\rm T}=0$. This solution is
also obtained from (\ref{25}). It is interesting that the $\omega$
of other solutions of \cite{17} are not reproduced by using
(\ref{25}). It is because these solutions are unstable attractors
and therefore
do not describe the late-time behavior of the system.\\ \\
{\bf F. $V=\frac{1}{2}m^2\p^2+ \frac{1}{2}g^2\p^2\s^2+
(M^2-\lambda\s^2)^2/(4\lambda)$}\\ \\
This potential has been investigated in \cite{14-0}. In this case,
it can be easily seen that the potential has a unique saddle point
at $(\p^*,\s^*)=(0,0)$. Therefore
  \be\label{26}
  \frac{V'}{V}=\frac{1}{V}\left( \frac{\partial V}{\partial \p}\p'
  +\frac{\partial V}{\partial \s}\s' \right)_{(\p,\s)=(0,0)} =0,
  \ee
which results in $\omega_0=-1$. This is the same value obtained in
\cite{14-0}, denoted there by de Sitter solution. Note that here
$x_\p=x_\s=0$ and $y=1$. So eq.(\ref{16}) leads to
  \be\label{27}
  \frac{x'}{x}\rightarrow -\frac{2y'}{y},
  \ee
which again becomes zero at $y'=0$ and $y=1$.
\\ \\
{\bf G. $V=V_1+V_2+\lambda\sqrt{V_1V_2}$, where $V_1=
V_{\p_0}e^{-\lambda_1\p}$ and $V_2= V_{\s_0}e^{-\lambda_2\s}$}\\
\\
This is the potential whose attractors have been obtained in
\cite{18}. This potential has no extremum point, but it is clear
that $\p^*\rightarrow\infty$ and $\s^*\rightarrow -\infty$. In
this case
  \begin{align}\label{28}
  \frac{V'}{V}=&\frac{1}{V}\left[ -\lambda_1(V_1+\frac{\lambda}{2}
  \sqrt{V_1V_2})\p'-\lambda_2(V_2+\frac{\lambda}{2}
  \sqrt{V_1V_2})\s'\right]_{(V_1,V_2)\rightarrow(0,\infty)}
  \nonumber\\ =&-\lambda_2\s'.
  \end{align}
So
   \be\label{29}
  \omega_0=-1+\frac{1}{3}\lambda_2\s'.
  \ee
The only stable attractor of this potential is the second solution
of Table 1 of \cite{18} with coordinates
$x_\p=0,\,x_\s=-\lambda_2/\sqrt{6}$ and $y^2=1+\lambda_2^2/6$.
These values result in
$\omega=-1-\lambda_2^2/3=-1+\lambda_2\s'/3$, which is same as
eq.(\ref{29}).

{\bf Acknowledgement:} This work was partially supported by the
"center of excellence in structure of matter" of the Department of
Physics.


\begin{thebibliography}{99}

\bibitem{1} A. G. Riess {\it et al.}, Astron. J. {\bf 116} (1998)
1009; S. Perlmutter {\it et al.}, Nature {\bf 391} (1998) 51; R.
A. Knop {\it et al.}, Astrophys. J. {\bf 598} (2003) 102.

\bibitem{2} S. W. Allen {\it et al.}, Mon. Not. Roy. Astron. Soc. {\bf 353} (2004)
457; M. Tegmark {\it et al.}, Phys. Rev. D {\bf 69} (2004) 103501;
A. C. Pope {\it et al.}, Astrophys. J. {\bf 607} (2004) 655.

\bibitem{3} C. Bennet {\it et al.}, Astrophys. J. (Suppl.) {\bf 148} (2003)
1; G. Hinshaw {\it et al.}, Astrophys. J. (Suppl.) {\bf 148}
(2003) 135; D. N. Spergel {\it et al.}, astro-ph/0603449.

\bibitem{4} S. Weinberg, Rev. Mod. Phys. {\bf 61} (1989) 1.

\bibitem{5} D. Huterer and A. Cooray,
Phys. Rev. D {\bf 71} (2005) 023506; S. Nesserisa and L.
Perivolaropoulos, Phys. Rev. D {\bf 72} (2005) 123519; U. Seljak,
A. Slosar and P. McDonald, JCAP {\bf 0610} (2006) 014.

\bibitem{6} A. R. Liddle P. Parson and J. D. Barrow, Phys. Rev. D {\bf 50}
(1994) 7222; R. R. Caldwell, R. Dave and P. J. Steinhardt, Phys.
Rev. Lett. {\bf 80} (1998) 1582; P. J. E. Peebles and A. Vilenkin,
Phys. Rev. D {\bf 59} (1999) 063505;  M. Doran and J. Jaeckel,
Phys. Rev. D {\bf 66} (2002) 043519; H. Ziaeepour, Phys. Rev. D.
{\bf 69}(2004) 063512; M. Garny, Phys. Rev. D {\bf 74} (2006)
043009; Xin Zhang, Phys. Lett. B {\bf 648} (2007) 1.

\bibitem{7} R. R. Caldwell, Phys. Lett. B {\bf 545}
(2002) 23; R. R. Caldwell, M. Kamionkowski and N. N. Weinberg,
Phys. Rev. Lett. {\bf 91} (2003) 071301; J. M. Cline, S. Y. Jeon
and G. D. Moore, Phys. Rev. D {\bf 70} (2004) 043543.

\bibitem{8}B. Feng, X. L. Wang and X. M.
Zhang, Phys. Lett. B {\bf 607} (2005) 35.

\bibitem{9} H. Mohseni Sadjadi and M. Alimohammadi, Phys. Rev. D {\bf
74} (2006) 043506; M. Alimohammadi and H. Mohseni Sadjadi, Phys.
Lett. B {\bf 648} (2007) 113.

\bibitem{10} L. P. Chimento, A. S. Jakubi, D. Pavon and W.
Zimdahl, Phys. Rev. D {\bf 67} (2003) 083513; H. Mohseni Sadjadi
and M. Alimohammadi, Phys. Rev. D {\bf 74} (2006) 103007.

\bibitem{11} E. J. Copeland, A. R. Liddle and D. Wands, Phys. Rev.
D {\bf 57} (1998) 4686; A. de la Macorra and G. Piccinelli, Phys.
Rev. D {\bf 61} (2000) 123503; S. C. C. Ng, N. J. Nunes and F.
Rosati, Phys. Rev. D {\bf 64} (2001) 083510; X. Li, Y. Zhao and C.
Sun, Class. Quant. Grav. {\bf 22} (2005) 3759; V. Faraoni, Class.
Quant. Grav. {\bf 22} (2005) 3235; A. D. Rendall, Class. Quant.
Grav. {\bf 22} (2005) 1655; M. Alimohammadi and H. Mohseni
Sadjadi, Phys. Rev. D {\bf 73} (2006) 083527.

\bibitem{12} J. Kujat, R. J. Scherrer and A. A. Sen, Phys. Rev. D {\bf 74}
(2006) 083501.

\bibitem{13} W. Zhao and Y. Zhang, Phys. Rev. D {\bf 73} (2006)
123509.

\bibitem{14-0} R. Lazkoz, G. Leon and I. Quiros,
Phys. Lett. B {\bf 649} (2007) 103.

\bibitem{14} P. J. Steinhardt, L. Wang and I. Zlatev,
Phys. Rev. D {\bf 59} (1999) 123504.

\bibitem{15} T. Chiba, Phys. Rev. D {\bf 73} (2006) 063501.

\bibitem{16} Z. K. Guo, Y. S. Piao, X. Zhang and Y. Z. Zhang,
Phys. Lett. B {\bf 608} (2005) 608.

\bibitem{161} Z. K. Guo, Y. S. Piao, X. Zhang and Y. Z. Zhang,
Phys. Rev. D {\bf 74} (2006) 127304.

\bibitem{17} R. Lazkoz and G. Leon, Phys. Lett. B {\bf 638} (2006) 303.

\bibitem{18} X. Zhang, H. Li, Y. S. Piao and X. Zhang, Mod. Phys.
Lett. A {\bf 21} (2006) 231.


\end{thebibliography}
\end{document}